\documentclass[10pt,twocolumn,aps,prl,superscriptaddress,reprint]{revtex4-1}

\usepackage[latin9]{inputenc}
\usepackage{bm}
\usepackage{amsmath}
\usepackage{graphicx}
\usepackage[unicode=true,
 bookmarks=false,
 breaklinks=false,pdfborder={0 0 1},colorlinks=false]
 {hyperref}
\usepackage{breakurl}

\usepackage{float}
\usepackage{threeparttable} 
\usepackage{multirow}

\makeatletter

\usepackage{xcolor}
\usepackage{bm}
\usepackage{url}

\makeatother

\newcommand{\be}{\begin{equation}}
\newcommand{\ee}{\end{equation}}

\newcommand{\bk}{{\bm{k}}}
\newcommand{\br}{{\bm{r}}}

\newcommand{\bq}{{\bm{q}}}

\newenvironment{sequation}{\begin{equation}\small}{\end{equation}}

\begin{document}

\title{On the charge and magnetic ordering in monolayer NbSe$_2$:\\ a first principles study}

\author{Feipeng Zheng}
\email[These coauthors contribute equally to this work.]{}
\affiliation{International Center for Quantum Materials, School of Physics, Peking
	University, Beijing 100871, China}
\affiliation{Siyuan Laboratory, Guangzhou Key Laboratory of Vacuum Coating Technologies and New Energy Materials, Department of Physics, Jinan University, Guangzhou 510632, China}

\author{Zhimou Zhou}
\email[These coauthors contribute equally to this work.]{}

\author{Xiaoqiang Liu} 
\affiliation{International Center for Quantum Materials, School of Physics, Peking
	University, Beijing 100871, China}
	
\author{Ji Feng}
\email{jfeng11@pku.edu.cn}
\affiliation{International Center for Quantum Materials, School of Physics, Peking
University, Beijing 100871, China}
\affiliation{Collaborative Innovation Center of Quantum Matter, Beijing 100871,
China}

\date{\today}
\begin{abstract}
Monolayer NbSe$_2$ has recently been shown to be a 2-dimensional superconductor, with a competing charge-density wave (CDW) order. This work investigates the electronic structure of monolayer NbSe$_2$ based on first principles calculations, focusing on charge and magnetic orders in connection to the superconductivity. It is found that decreased screening in the monolayer NbSe$_2$ with a perfect lattice exhibits magnetic instability, which is removed by the formation of CDW. Two energetically competitive but distinct $3\times3$ CDW structures are revealed computationally, which have a significant impact on the Fermi surface. The relations of the potential CDW phases with experimental structure and the coexisting superconductivity are discussed.

\end{abstract}
\maketitle

Layered transition metal dichalcogenides (TMDCs) MX$_2$, where M is a transition metal and X is a chalcogen, are a remarkable class of materials displaying a multitude of correlation effects, ranging from CDW, magnetic ordering to superconductivity. In recent years, TMDCs have become available in monolayer forms~\cite{Xi2015,Sarkar2015,Wang2012,Tang2017,Li2015}. In particular, monolayer niobium diselenide, NbSe$_2$, has an extraordinarily rich phase diagram with respect to temperature. In the bulk form, NbSe$_2$ is one of the first materials found to host coexisting CDW and superconducting orders~\cite{Revolinsky1965,Moncton1975}. In the monolayer limit, superconductivity again coexists with the CDW order, with a superconducting $T_c$ = 1.9 K~\cite{Ugeda2015} compared with the bulk value 7 K. A commensurate charge-density wave (CDW) transition was found  at 145 K~\cite{Xi2015}, with CDW vector $\boldsymbol{q} = \frac{1}{3}\bm a^*$, corresponding to a structural reconstruction within a 3$\times$3 supercell. In the monolayer limit, on the other hand, it is well known that screening is significantly reduced compared to the bulk counterpart, leading sometimes to dramatically enhanced electronic correlation. Indeed, previous first principles calculations suggest possible antiferromagnetic order in monolayer NbSe$_2$ in the absence of CDW order~\cite{Xu2014, Zhou2012}. The above observations indicate that the correlations arising from lattice and interaction are both important in monolayer NbSe$_2$. 

On account of such multi-correlated nature, the interplay of different correlation effects, and therefore possible phases of monolayer NbSe$_2$, are yet to be clarified, which requires treating different instabilities on the same footing. One aim of the present paper, therefore, is  to analyze the equilibrium atom arrangement in the CDW phase of monolayer NbSe$_2$, and at the same time, the competition between the CDW and magnetic instabilities. It is revealed that the formation of CDW phase eventually suppresses the magnetic instability, providing clarification to the vagary regarding the absence of magnetic order in monolayer NbSe$_2$. A second objective of this paper is to understand the electronic structure of monolayer NbSe$_2$, in the eventual low-temperature CDW phase. A Brillouin zone unfolding scheme is devised to compare the Fermi surface of the CDW phase to that of the symmetric lattice phase. Whereas the monolayer NbSe$_2$ with symmetric lattice has three Fermi circles, at $\mathbf{\Gamma}$, $\mathbf{K}$ and $\mathbf{K'}$, respectively, the CDW order leaves them partially or fully gapped. The extensive obliteration of the computed Fermi surface leads to an explanation for the disappearance of the magnetic instability, and potentially, for the lower-than-bulk superconducting $T_c$ for monolayer NbSe$_2$.

\begin{figure}[H]
	\centering
	\includegraphics[width=76 mm]{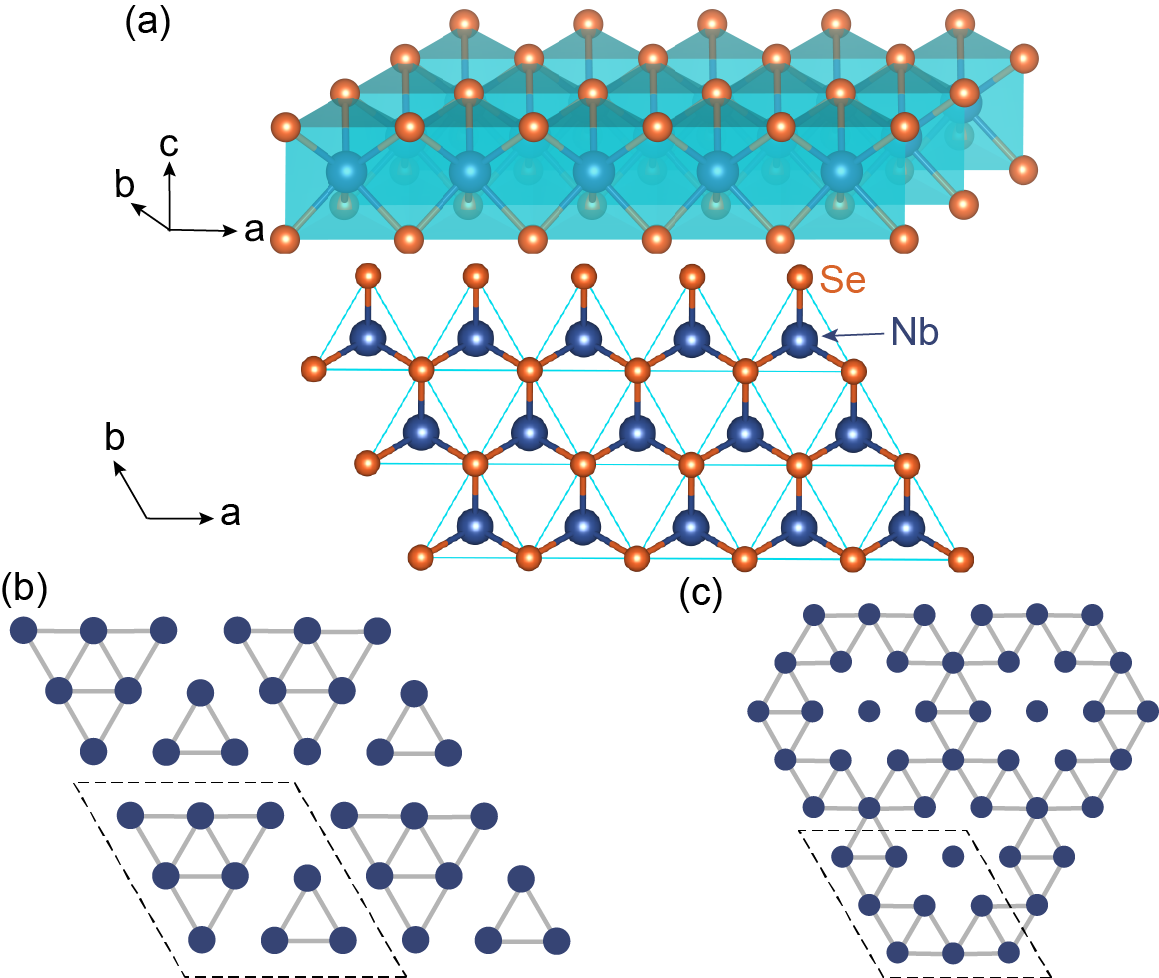} \\
	\caption{ (a) Top and side views of crystal structure of monolayer NbSe$_2$ without CDW. Distributions of the Nb atoms in the triangle CDW phase (b) and the star CDW phase (c).  The solid gray bonds between the adjacent Nb atoms indicate that their distances were shortened after CDW transition. The rhombuses bounded by dashed lines in (b) and (c) denote 3$\times$3 CDW supercells.}
	\label{fig:str_pho}
\end{figure}

The structure of the monolayer NbSe$_2$ without CDW, isolated from the bulk phase $2H$-NbSe$_2$, forms a 2-dimensional hexagonal lattice in the 2-dimensional space group $P\bar 6m2$, as shown in Fig.~\ref{fig:str_pho}(a). It is composed of three layers of atoms, with a Nb layer sandwiched between Se layers. Each Nb atom sits inside a trigonal prismatic cage formed by six nearest-neighbor Se atoms. Nb atoms form a perfect hexagonal closest packing  structure, with the  shortest  Nb-Nb separation  $a = 3.474$ \AA  ~according to our calculations~\cite{SM}. 
This non-CDW phase of monolayer NbSe$_2$ is found to be unstable below $T_{\textrm{CDW}}$~\cite{Calandra2009} and exhibits pronounced soft phonon mode around  $\boldsymbol{q} = \tfrac{1}{3}\bm a^{*}$ in our calculations, indicating a strong structural instability to the formation of a 3$\times$3 supercell.

Density-functional theory (DFT) calculations were performed, within the generalized gradient approximation (GGA), parameterized by Perdew, Burke, and Ernzerhof (PBE) to investigate the crystal structure, electronic structure and lattice dynamics of monolayer NbSe$_2$ in non-CDW and CDW phase \cite{Kresse1996,Perdew1996}. The Kohn-Sham valence states were expanded in the plane wave basis set with a kinetic energy truncation at 530 eV. An 18$\times$18$\times$1 and a 6$\times$6$\times$1  $\mathbf{k}$-grid centered at $\Gamma$ point were chosen for the sampling of the  Brillouin zones of the non-CDW and CDW phase respectively. The equilibrium crystal structure was determined by a conjugated-gradient relaxation of ionic positions, until the Hellmann-Feynman force on each atom was less than 0.005 \AA /eV and a zero-stress tensor was obtained. Crystal structure of the NbSe$_2$ in the CDW state was obtained with multiple structural optimizations with a 3$\times$3 supercell. Before each optimization, small random displacements were added to each atom to remove all discrete symmetries. 

When the structure of monolayer  NbSe$_2$ is relaxed in a $3\times 3$ supercell, the CDW indeed forms, which can be characterized by a reconstruction of the Nb atomic layer, with  concomitant displacements of Se atoms. 
Besides, we also examine  the lattice stabilities of the three reconstructed CDW phases under different in-plane lattice constants and using different types of exchange-correlation functionals, which give rise to similar results as described in the supplemental material~\cite{SM}. 
Using the fully relaxed in-plane lattice constant $a = 3.474$~\AA, 
two $3\times3$ CDW reconstructions are found to be important, referred to herein as triangle and star phases (see supplemental material for details~\cite{SM}).  Fig.~\ref{fig:str_pho}(b) displays  the triangle phase, where the Nb atoms are grouped into large and small  triangular clusters, consisting of six and three Nb atoms respectively within a 3$\times$3  supercell.  
The star phase is characterized by overlapping star-shaped clusters, as displayed in Fig.~\ref*{fig:str_pho}(c), which is similar to bulk $2H$-NbSe$_2$ in CDW state~\cite{Malliakas2013,Silva2016}.

The topography of monolayer NbSe$_2$ in the CDW phase, which may be revealed by scanning-tunneling microscopy (STM), is simulated. We compute the local electronic density of states, averaged over an energy window below the Fermi level
\begin{sequation}
	\rho_{\text{avg}}(\br) = \frac{A}{\Delta} \sum_{n}\int_{\mu-\Delta}^\mu d\varepsilon \int\frac{\text{d}^2k}{(2\pi)^2}
	\psi_{n\bk}^*(\br)\psi_{n\bk}(\br)\delta(\varepsilon-\varepsilon_{n\bk}),
\end{sequation}
where $\psi_{n\bk}(\br)$ is the Kohn-Sham wavefunction of the $n$th band at wavevector $\bk$, $\mu$ the Fermi energy, $\Delta$ the width of the energy window, and $A$ the area of a supercell. The energy window is set to $\Delta = -~4$ meV. The simulated STM topography, $z=z(x,y)$, is subsequently determined by the implicit equation, $\rho_{\text{avg}}(x,y,z)=\rho_0$, where the iso-density value $\rho = $ 6.3$\times10^{-4}$ $e/$\AA$^{2}$~is used to best match the experimental STM topography~\cite{Xi2015}. Since the tunneling current  is proportional to local density of states, $z(x,y)$  can be interpreted semiquantitively as the topography revealed in a constant-current STM topographical scan.

\begin{figure}[H]
	\centering
	\includegraphics[width=76mm]{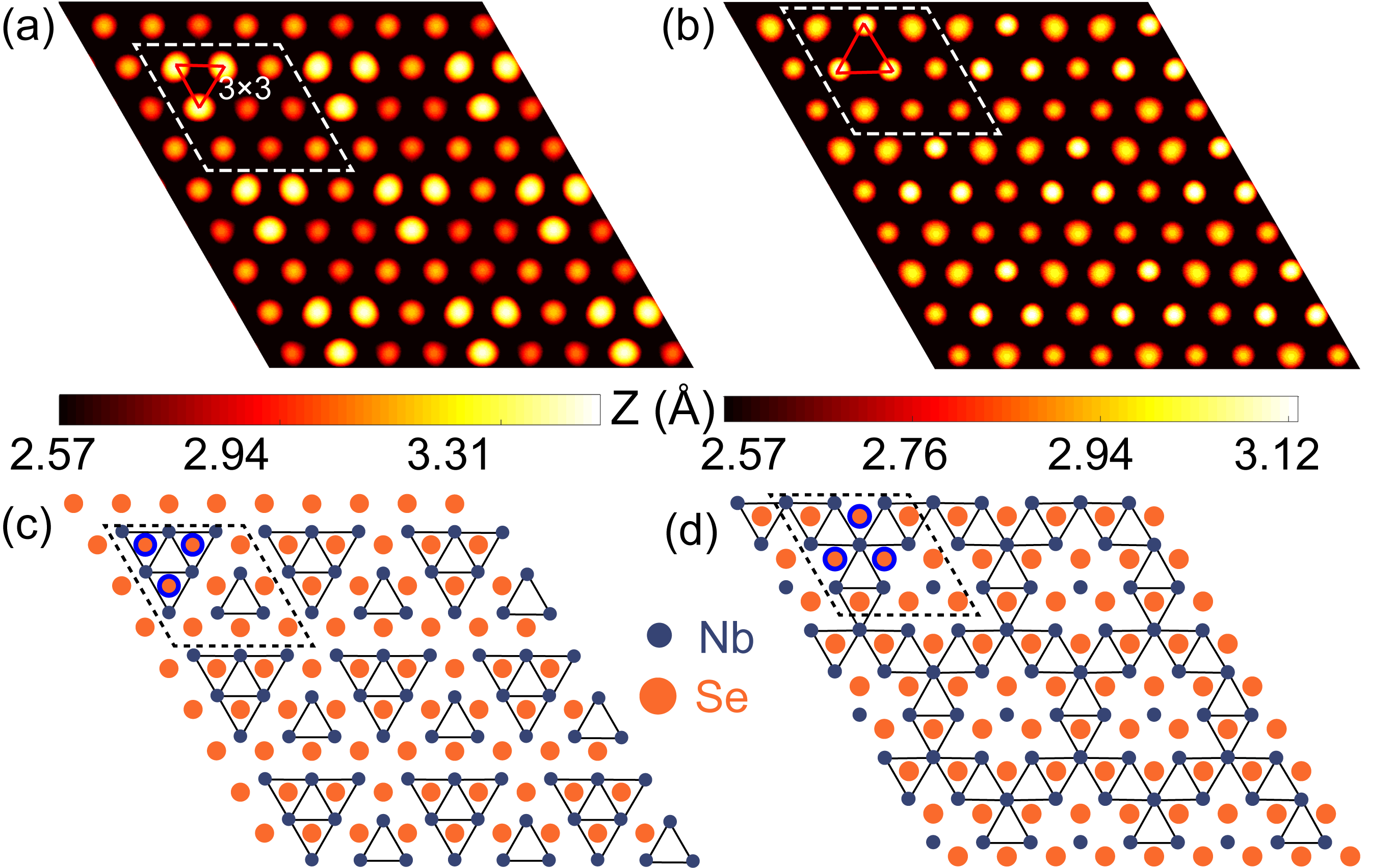} \\
	\caption{Computed STM topographies of monolayer NbSe$_2$ in the triangle (a) and star (b) CDW phase. The colors in the map represent the height function $z(\mathbf{x}, \mathbf{y})$ as mentioned in the main text, where $\mathbf{x}$  and $\mathbf{y}$ are in-plane coordinates. The height of Nb layer was set to be $z$ = 0. The red  triangles in (a) and (b) indicate the dominant features of the corresponding topography within a 3$\times$3 CDW supercell bounded by white dashed lines. Top view of the configurations of the triangle (c) and star (d) CDW phase.  The Se atoms with blue circles in (c) and (d) are corresponding to the dominant STM signals marked by the red triangles in (a) and (b) respectively as mentioned above.
	}
	\label{fig:STM}
\end{figure}

The simulated STM topographies of the triangle and star CDW phases are displayed in Fig.~\ref{fig:STM}. The STM topography is dominated by top Se atoms (see Fig.~S2~\cite{SM}), despite that the electronic states around the Fermi level arise mostly from the $d$-orbitals of Nb atoms. Furthermore, the dominant STM feature of the triangle  phase  is contributed by the three Se atoms on the top of the large triangular cluster, shown as a red triangle as displayed in Figs.~\ref{fig:STM}(a) and \ref{fig:STM}(c) (marked with three blue circles).  
Similarly, the star phase can be also characterized by similar red triangles as displayed in Fig.~\ref{fig:STM}(b), attributable to the three top-layer Se atoms, indicated by blue circles in Fig.~\ref{fig:STM}(d).
Comparing with the experimental STM topography shown in Fig.~1(e) of the reference~\cite{Ugeda2015}, which shows an array of small triangles, we find that the simulated STM topographies of triangle and star phases are both consistent with the experiment. 
Owing to the similarity of STM patterns and energies (see supplemental material~\cite{SM}) of the triangle and star phases, we are unable to suggest which CDW is more likely to occur in the actual monolayer NbSe$_{2}$.


In the monolayer limit, the screening is substantially reduced compared to bulk material, giving rise to stronger interaction between electrons. It is then expected that monolayer material is more prone to developing magnetic order even with a non-magnetic bulk counterpart. Previous first principles investigations \cite{Xu2014,Zhou2012} indicate that the most stable state of monolayer  NbSe$_2$ without CDW order is antiferromagnetic in a $4\times1$ supercell. The issue, however, remains with whether the magnetic order coexists or competes with the CDW order,  both of which are known to modify the Fermi surface and consequently the susceptibility. Therefore, in order to quantify the magnetic instability, we analyse the spin susceptibility of monolayer NbSe$_2$, obtained from tight-binding models based upon Wannier functions extracted from the Kohn-Sham band structures~\cite{Mostofi2014}.

The general form of noninteracting susceptibilities reads~\cite{Kubo2007}:
\begin{multline}
	\chi_{st}^{pq}\left(\boldsymbol{q},\omega\right)=-\frac{1}{N}\sum_{\boldsymbol{k},\mu\nu}\left[f\left(E_{\nu}\left(\boldsymbol{k+q}\right)\right)-f\left(E_{\mu}\left(\boldsymbol{k}\right)\right)\right]\\
	\cdotp\frac{a_{\mu}^{s}\left(\boldsymbol{k}\right)a_{\mu}^{p*}\left(\boldsymbol{k}\right)a_{\nu}^{q}\left(\boldsymbol{k+q}\right)a_{\nu}^{t*}\left(\boldsymbol{k+q}\right)}{\omega+E_{\nu}\left(\boldsymbol{k+q}\right)-E_{\mu}\left(\boldsymbol{k}\right)+i0^{+}}.
\end{multline}
where $s,t,p,q$ denoting orbital index, $N$ is the number of lattice
sites, and $a_{\mu}^{s}\left(\boldsymbol{k}\right)=\left\langle s|\mu\boldsymbol{k}\right\rangle $,
is the amplitude of the $\mu$th band at crystal momentum $\bk$ on the $s$th Wannier orbital, obtained by
diagonalizing the tight-binding Hamiltonian.
The static one-loop spin susceptibility is given by $\chi_{S}\left(\boldsymbol{q}\right)=\frac{1}{2}\sum_{s,p}\chi_{ss}^{pp}\left(\boldsymbol{q},\omega=0\right)$.
In our non-spin-polarized DFT calculations of $3\times3$ supercell of
monolayer NbSe$_2$,  with and without (triangle or star) CDW and in-between, the manifold of 9 bands
around the Fermi level are predominantly formed by Nb $d_{z^{2}}$
orbitals, which are well separated from all other bands. Therefore, these bands are used to construct 9-band tight-binding Hamiltonians.

\begin{figure}[H]
	\centering
	\includegraphics[width=74mm]{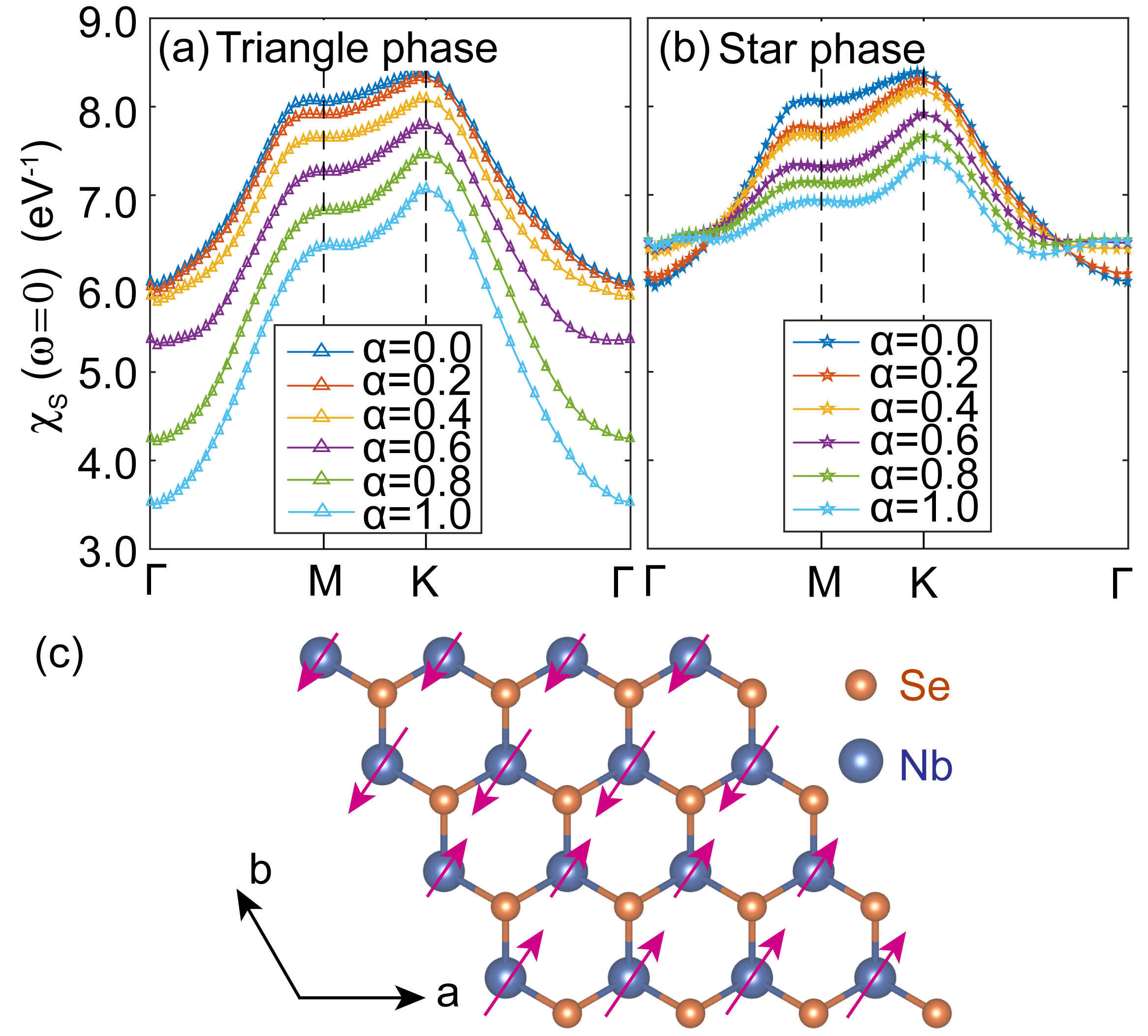} \\
	\caption{ Calculated noninteracting static spin susceptibilities $\chi_{S}\left(\boldsymbol{q},\omega=0\right)$ in $3\times3$ supercells for different $\alpha$ when the CDW distortions are triangle-like(a) and star-like(b). (c) The calculated spin configurations in antiferromagnetic non-CDW phase. The high magnetic moment($\pm0.41\mu_B$ per Nb atom) and low magnetic moment($\pm0.12\mu_B$ per Nb atom) are denoted by long and short arrows, respectively.
	}
	\label{fig:sus}
\end{figure}

In order to assess how the magnetic instability evolves as the CDW is developing, we calculated the spin susceptibilities $\chi_S(\bq)$ of a series structures interpolated between the non-CDW and CDW phases. 
Here, the atom posititions of an interpolated supercell is given by $\boldsymbol{r}(\alpha)=\left(1-\alpha\right)\boldsymbol{r}_{0}+\alpha\boldsymbol{r}_{c}$, where $\alpha$ corresponds to the amplitude of CDW distortions, and $\boldsymbol{r}_{0}$ and $\boldsymbol{r}_{c}$  are the positions of atoms in state without and with CDW distortions, respectively. 
Figs.~\ref{fig:sus}(a)-\ref{fig:sus}(b) show the calculated  $\chi_{S} \left(\boldsymbol{q},\omega=0\right)$ for different $\alpha$ in $3\times3$ supercells. When $\alpha=0$, corresponding to non-CDW phase, the spin susceptibility peaks at around $\mathbf{M}$ and $\mathbf{K}$, which indicates magnetic instability, to the formation of a magnetic supercell. This is consistent with a spin-polarized calculation, leading to a magnetic configuration as shown in Fig.~\ref{fig:sus}(c)~\cite{Xu2014}. It can be clearly seen that with the increase of $\alpha$, the spin susceptibilities generally decrease for  both of the triangle and star CDW, suggesting that as the CDW order forms the magnetic order may eventually disappear. However, the spin susceptibilities of star phase are slightly larger than that of the triangle phase, which indicates that their magnetic behaviors may be different when CDW distortions set in.  

Our spin spiral calculations (see supplemental material for details~\cite{SM}) suggest that the non-CDW phase of monolayer NbSe$_2$ possesses an antiferromagnetic magnetic ground state as depicted in Fig.~\ref{fig:sus}(c), which is in agreement with previous predictions~\cite{Xu2014}. According to our spin susceptibility calculations above, this magnetic order may eventually be suppressed or even killed by CDW. As a confirmation, we computed the total energy of a $12\times 3$ supercell to include in both magnetic and CDW orders. We used the interpolation method described above to get a series structures with different values of $\alpha$  and carried out both non-spin-polarized and spin-polarized calculations for each kind of CDW.

\begin{figure}[H]
	\centering
	\includegraphics[width=74mm]{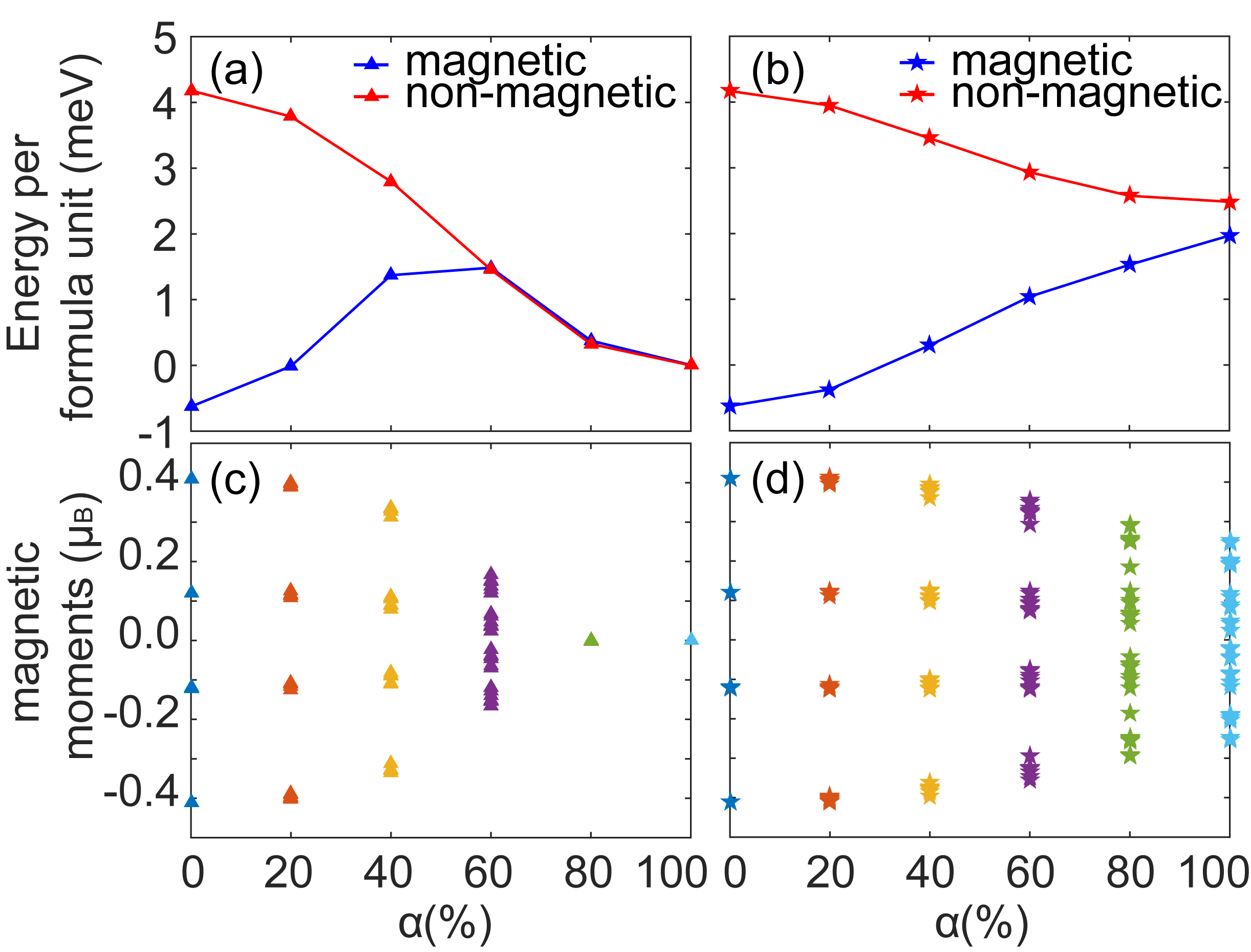} \\
	\caption{(a) Calculated total energy of $12\times3$ supercells in magnetic (blue) and non-magnetic (red) calculations for triangle CDW (a) and star CDW (b), respectively. The energy of the triangle CDW phase is set to be zero. Computed magnetic moments of Nb atoms in $12\times3$ supercells at various values of $\alpha$ for triangle CDW (c) and star CDW (d), respectively.}
	\label{fig:Emk-nmk}
\end{figure}

Figs.~\ref{fig:Emk-nmk}(a)-\ref{fig:Emk-nmk}(b) show the calculated total energies, and Figs.~\ref{fig:Emk-nmk}(c)-\ref{fig:Emk-nmk}(d) show the calculated magnetic moments of Nb atoms at various values of $\alpha$ for the triangle and star CDW, respectively.  It is found that for both kinds of CDW, when an antiferomagnetic order is imposed, the total energy increases with the increase of $\alpha$. Meanwhile, the magnetic moments are suppressed, which indicates a competition between CDW and SDW (spin-density wave).
However, things are slightly different for triangle and star CDW. For triangle CDW, it is found that when $\alpha < 0.6$, magnetic phase, which has a collinear magnetic order as shown in Fig.~\ref{fig:sus}(c), has lower energy than the non-magnetic phase as displayed in Fig.~\ref{fig:Emk-nmk}(a). As $\alpha$ increases, the difference of total energy between magnetic and non-magnetic calculations decreases gradually. At about $\alpha=0.6$, the energetic advantage of magnetic phase disappears, beyond which the CDW order suppresses the magnetic order completely, which can be verified from the zero magnetic moments when $\alpha > 0.6$, as shown in Fig.~\ref{fig:Emk-nmk}(c). 

For star CDW, although as $\alpha$ increases, the difference of total energies between magnetic and non-magnetic calculations decreases gradually, the magnetic order remains, as shown in Figs.~\ref{fig:Emk-nmk}(b) and \ref{fig:Emk-nmk}(d). The general trend of the magnetic phases' energies in the two kinds of CDW is also consistent with our previous spin susceptibilities calculations. 
Base on the results above, we then conclude that without CDW, monolayer NbSe$_{2}$ favors antiferromagnetic order in a $4\times1$ supercell. When the CDW order sets in, the magnetic order is suppressed by the triangle CDW whereas the star CDW may coexist with magnetism.


To investigate the influence of CDW on electronic structure of monolayer NbSe$_2$, band structures and Fermi surfaces of the non-CDW and CDW phases were computed using a 3$\times$3 supercell and unfolded to the (larger) primitive Brillouin zone of the non-CDW NbSe$_2$, by a weight factor measuring the tranformation of wavefunctions under primitive lattice translations (see supplemental material for details~\cite{SM}).

\begin{figure}[h]
	\centering
	\includegraphics[width=76mm]{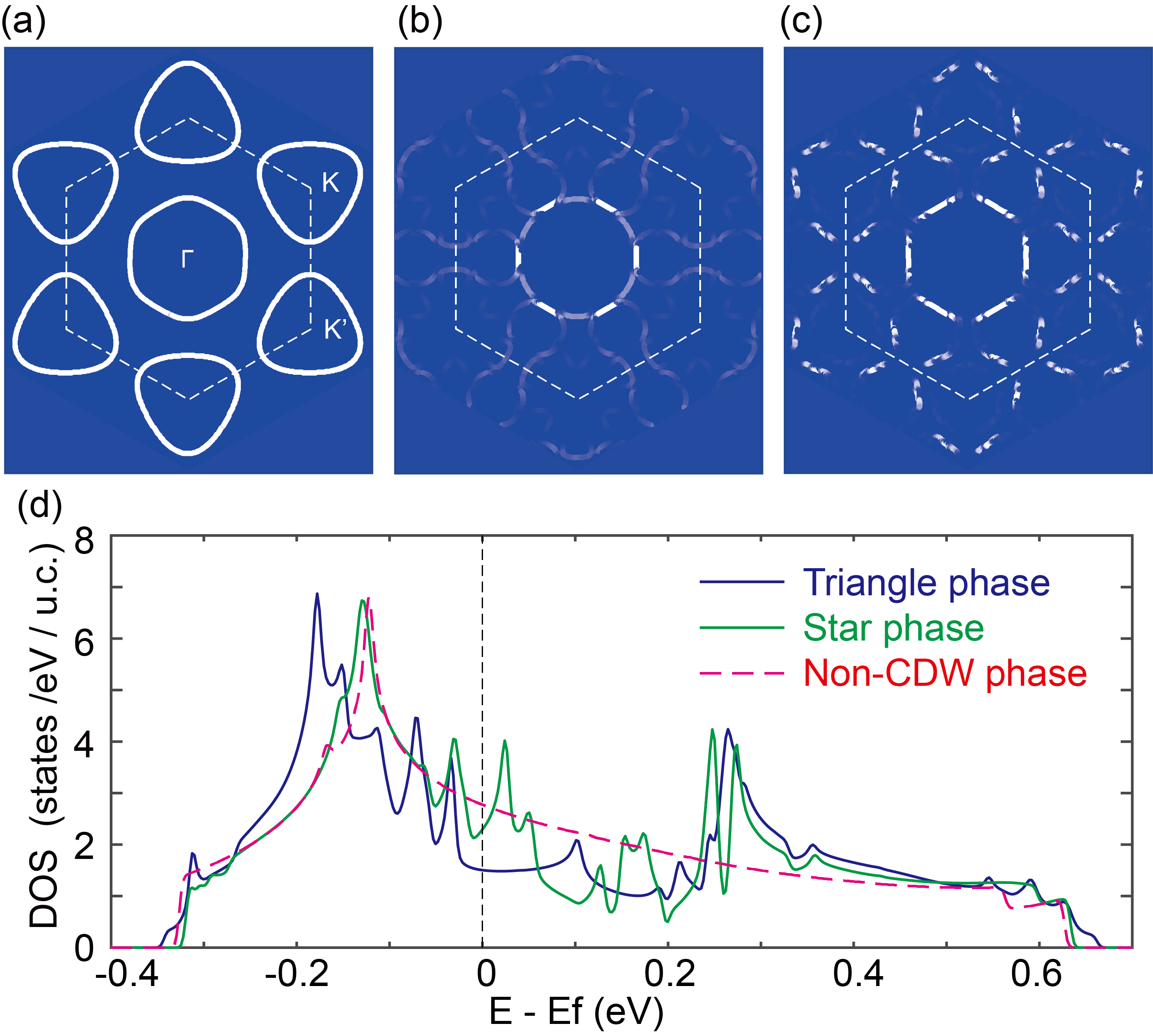} \\
	\caption{Unfolded Fermi surfaces of the non-CDW phase (a), triangle CDW phase (b), and star CDW phase (c) in primitive Brillouin zone of the non-CDW phase, respectively. The brightness of the dots in (a)-(c) denotes corresponding unfolding weights $W_{\mathbf{k}J}$. The hexagon (white dashed line) is the boundary of primitive Brillouin zone. 
	(d) Computed electronic density of states of the triangle, star and non-CDW phases.}
	
	\label{fig:unfold}
\end{figure}

As shown in Fig.~\ref{fig:unfold}(a), the Fermi surface of the non-CDW phase consists of three inequivalent circles (white lines), centered at $\mathbf{\Gamma}$, $\mathbf{K}$ and $\mathbf{K'}$  points, respectively. After CDW transition, they are partially or fully gapped due to the lattice distortion. 
Fig.~\ref{fig:unfold} displays the unfolded Fermi surfaces of the non-CDW phase (a), triangle CDW phase (b), and star CDW phase (c), respectively. It is evident that the triangle phase develops more extensive gapping on the Fermi surface compared to the star phase. Both of the triangle and star CDW phases show significant partial gapping along the Fermi surface encircling $\mathbf{\Gamma}$, which appears to be more extensive for the triangle phase.
Remarkably, while the $\mathbf{K}$-, $\mathbf{K'}$-centered Fermi circles are only partially gapped in the star phase, they are almost completely obliterated in the triangle phase.
The computed electronic density of states also show the consistent results. As displayed in Fig.~\ref{fig:unfold}(d), the density of states of star phase at the Fermi level exhibits slight decrease, compared with that of the non-CDW phase. In contrast, the triangle phase shows a more pronounced reduction of the Fermi density of states, corresponding to the more extensive CDW gap on the Fermi surface.

The modification of the Fermi surface also provides a microscopic insight into the interplay between CDW, SDW and superconductivity. In the non-CDW phase with the intact Fermi surface, there exist electronic instabilities which can potentially trigger CDW or SDW. In the presence of the CDW along with the modification of Fermi surface, the SDW instability was subdued as discussed above. It  suggests  that the electronic states triggering  SDW are also responsible for the CDW instability, which are removed from Fermi surface when the CDW forms.

In summary, our calculations reveal that the formation of CDW in monolayer NbSe$_2$ suppresses the magnetic instability of the initiating lattice. Two possible CDW phases are identified computationally, which have similar STM topographical features in our simulation. The formation of CDW gap on the Fermi surface, as visualized from our band unfolding scheme, should also have an important impact on  superconductivity. It is expected that the extent of CDW gap is inversely related to the superconducting $T_c$. The Fermi gapping is substantially more extensive in the triangle phase, in comparison with the bulk-like star phase. In view of the substantially lower $T_c$ of the monolayer NbSe$_2$ compared to the bulk and our computational observation that the star CDW phase remains magnetic, these results potentially favor the hypothesis that the triangle CDW might actually form.

\begin{acknowledgements} This work was supported by National Natural Science Foundation of China (Grant No. 11725415), and Ministry of Science and  Technology of the People's Republic of China (Grant No. 2016YFA0301004). Part of the calculations were performed on the Tianhe-I Supercomputer System. 
\end{acknowledgements}

\makeatother

\renewcommand{\thepage}{S\arabic{page}}  
\renewcommand{\thesection}{S\arabic{section}}   
\renewcommand{\thetable}{S\arabic{table}}   
\renewcommand{\thefigure}{S\arabic{figure}}

\onecolumngrid

\newpage
\begin{center}
\large \textbf{SUPPLEMENTAL MATERIALS}
\end{center}

\section{Geometry structures of the  CDW phases and their lattice stabilities}

\begin{figure}[H]
	\centering
	\includegraphics[width=150mm]{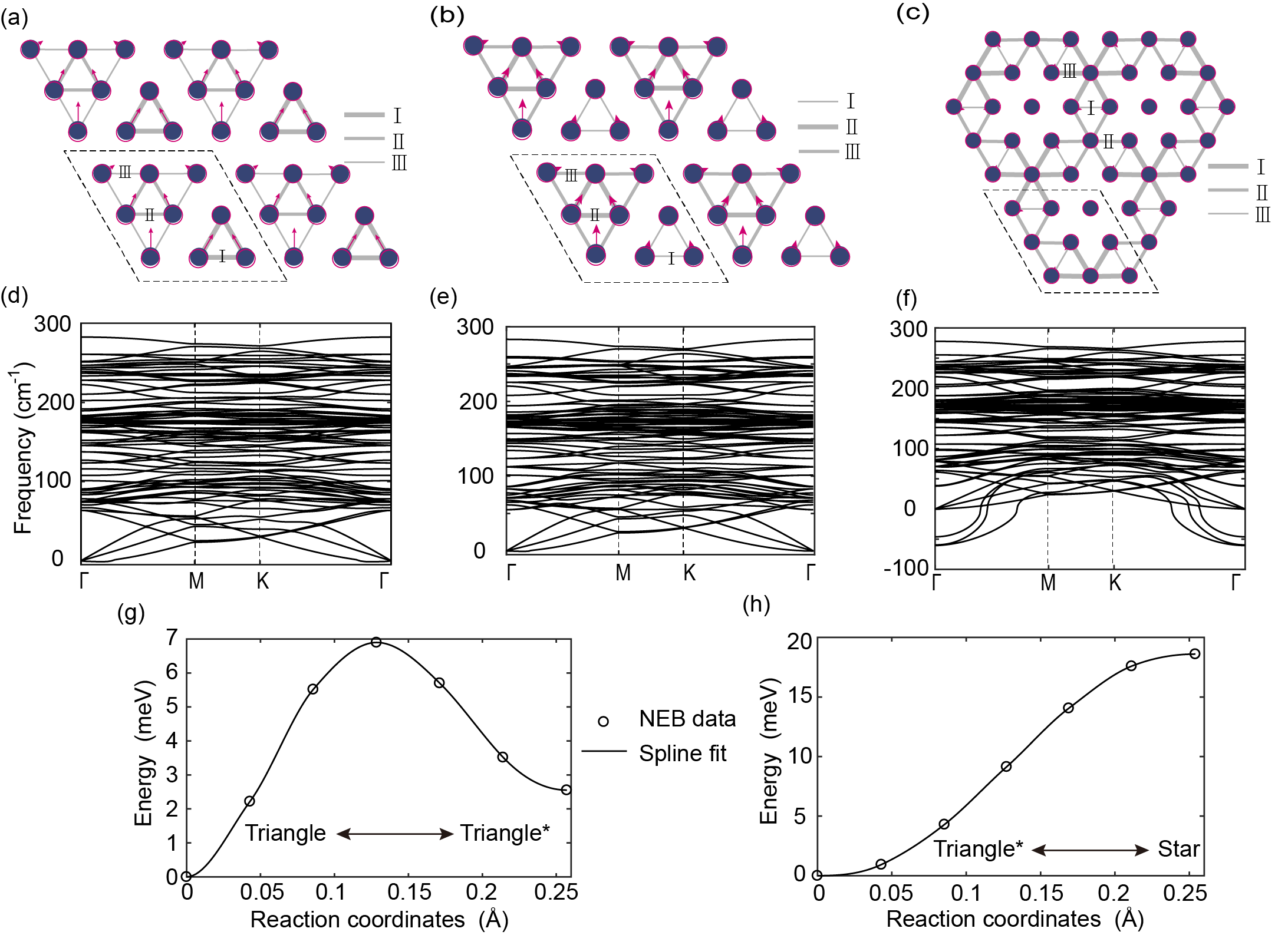} \\
	\caption{ Configurations of the Nb layer in the triangle  (a), triangle* (b) and star (c) CDW phases. The  magenta circles donate perfect hcp nets. The displacements of Nb atoms from perfect hcp nets due to CDW are illustrated using magenta arrows. The solid gray bonds between the adjacent Nb atoms indicate that their distances are shortened after CDW transition.  The gray bonds in (a)-(c)  can be classified as three types displayed with different thicknesses according to the distance of connected Nb atoms. The thicker a bond is, the smaller the distance is. Phonon dispersions of the the triangle (d),  triangle* (e) and star (f) CDW phases. Related total energies along reaction coordinates between the triangle and  triangle* phase (g), the triangle* and star phase (h).}
	\label{figs:str_pho_SI}
\end{figure}

This part mainly reports the details of the geometry structures and lattice stabilities of the triangle, triangle* and star CDW phases obtained  at $a = 3.474$~\AA~as  supplemental information of the main text. Three types of reconstructed structures were obtained. Two of the structures are characterized by triangular clusters (the triangle and triangle*  phase).  Fig.~\ref{figs:str_pho_SI}(a) displays  the triangle phase, where the Nb atoms are grouped into large and small  triangular clusters, consisting of six and three Nb atoms respectively within a 3$\times$3  supercell. The shorten Nb-Nb pairs after CDW transition can be classified as three types according to their distances marked by gray bonds with different thicknesses as illustrated in Fig.~\ref*{figs:str_pho_SI}(a) and the corresponding distances were listed in Table \ref*{table_s1}. This structure exbihits the lowest total energy.

The distributions of Nb atoms between the phases of triangle and triangle* are very similar as shown in Figs.~\ref{figs:str_pho_SI}(a)-\ref{figs:str_pho_SI}(b). The only difference  lies in the distance of the bond \uppercase\expandafter{\romannumeral1} as listed in Table \ref*{table_s1}. In detail, among the shorten pairs of Nb-Nb, the bond  \uppercase\expandafter{\romannumeral1} are the shortest  in triangle phase  but the largest in triangle* phase, which exhibits a slightly higher total energy as listed in Table \ref*{table_s1}. Interestingly,  both of the two phases are likely to be local minimums of the potential  energy surface according to the simulated results of  lattice dynamics [Figs.~\ref{figs:str_pho_SI}(d)-\ref{figs:str_pho_SI}(e)] and nudged elastic band (NEB)~\cite{Jonsson1998,Kresse1996} [Fig.~\ref{figs:str_pho_SI}(g)]. As a further confirmation, phonon spectrum was computed~\cite{Togo2015,Kresse1995,Alfe2001}  to ensure the obtained structures are free of soft phonon modes. In detail, as shown in Figs.~\ref{figs:str_pho_SI}(d)-\ref{figs:str_pho_SI}(e), the soft mode presenting in the non-CDW phase is completely removed in both  triangle and triangle* phases. Furthermore, the transition energy barrier between the two phases was computed. As displayed in  Fig.~\ref{figs:str_pho_SI}(g), the energy barrier is only 7 meV, indicating that the structural transition inbetween is rather easy.

Another reconstructed structure, star phase, was displayed in Fig.~\ref*{figs:str_pho_SI}(c). The star-shape clusters are grouped into two alternative units bounded by  \uppercase\expandafter{\romannumeral1} type of Nb-Nb as shown in Fig.~\ref{figs:str_pho_SI}(c).  The Nb layer can be viewed as composed of two sublattices owing to the modulation on Nb-Nb nearest-neighbour bonds. A continuous net of Nb forms a honeycomb lattice, with each of the bipartite sites decorated by a hexagonal cluster. The other sublattice is composed of an isolated Nb. It is easy to find that the displacements of Nb atoms in star  phase are much smaller than that in the triangle  phase by comparing the lengths of the magenta arrows in Fig.~\ref{figs:str_pho_SI}(c) with  Figs.~\ref{figs:str_pho_SI}(a)-\ref{figs:str_pho_SI}(b).  In contrast to the triangle and triangle* phases, the star phase is likely to be a metastable structure as shown in Figs.~\ref{figs:str_pho_SI}(f) and \ref{figs:str_pho_SI}(h).

\begin{table}[H]
	\begin{center}
		\label{table_s1}
		\caption{Nb-Nb distances (\AA) and related total energies (meV per chemical formula) of the non-CDW phase and the three CDW phases regrading the triangle, triangle* and  star  phase at the fully relaxed in-plane lattice constant 3.474 \AA~using PBE. The pairs  of  Nb-Nb  indexed  \uppercase\expandafter{\romannumeral1} -  \uppercase\expandafter{\romannumeral3} for the three CDW phases were shown in Figs.~\ref{figs:str_pho_SI}(a)-\ref{figs:str_pho_SI}(c).  }
		~\\
		\begin{tabular}{ccccccccccccc}
			\hline
			\hline
			Configuration ~~ &   Nb-Nb \uppercase\expandafter{\romannumeral1} (\AA)  ~~    &   Nb-Nb \uppercase\expandafter{\romannumeral2} (\AA)  ~~&  Nb-Nb \uppercase\expandafter{\romannumeral3} (\AA)~~& E (meV/ C.F.) \\
			\hline 		
			Triangle        &    3.33      &  3.37     & 3.41    &  0      \\
			\hline 
			Triangle*       &     3.42      &  3.38    &  3.40   &   0.3        \\
			\hline 
			Star               &   3.36       &  3.43     & 3.46   &  2.4    \\
			\hline 
			Non-CDW       &      3.47       & 3.47      & 3.47    & 3.7   \\
			\hline 
		\end{tabular}
		
	\end{center}
\end{table}

\begin{table}[H]
      \begin{center}
			\label{tabs:str_pho}
			\caption{ Related total energies of non-CDW  and the three CDW phases under different lattice constants computed using LDA and PBE.}
			~\\
			\begin{tabular}{ccccccccccc}
				\hline
				\hline
				Lattice  constant (\AA)   ~~    &  Functional              ~~~~~            &  Phase ~~~~~  &  E (meV/ C.F.) \\
				\hline
				\multirow{4}{*}{3.47$^{a}$}& 	\multirow{4}{*}{PBE} ~~~ &  Triangle    ~~~~    & 0  \\   
				&                                     &  Triangle*     ~~~~     &  0.3  \\	  
				&                                     &  Star          ~~~~       & 2.4  \\
				&                                     &  Non-CDW    ~~~~    & 3.7  \\     	 
				\hline       
				\multirow{4}{*}{3.399$^{a}$}& 	\multirow{4}{*}{LDA}~~~ &  Triangle   ~~~~        & 0.7 \\
				&                                     &  Triangle*     ~~~~    &   0  \\
				&                                     &  Star            ~~~~     &  2.9  \\	
				&                                     &  Non-CDW    ~~~~   &  3.3  \\	
				\hline        
				
				\multirow{8}{*}{3.458$^{a}$}   & 	\multirow{4}{*}{PBE}~~~  &  Triangle   ~~~~        & 0  \\   	 
				&                                     &  Triangle*     ~~~~     &  0.1  \\	  
				&                                     &  Star             ~~~~    & 2.4  \\
				&                                     &  Non-CDW   ~~~~     & 3.5  \\   
				\cline{2-4} 	      
				& 	\multirow{4}{*}{LDA} ~~~ &  Triangle    ~~~~       & 0.05  \\   
				&                                     &  Triangle*     ~~~~     &  0  \\	  
				&                                     &  Star              ~~~~   & 1.8  \\
				&                                     &  Non-CDW    ~~~~   & 3.4  \\     	 
				
				\hline 
			\end{tabular}
			
			\begin{tablenotes}
				\item[a] The lattice constants 3.399 and 3.474 \AA~are obtained by  fully relaxation of the lattices  using LDA and PBE respectively. 
				3.458  \AA~is the lattice constant of bulk $2H$-NbSe$_2$~\cite{Malliakas2013}. 				
			\end{tablenotes}
	\end{center}
\end{table}

We carefully examine the lattice stabilities of the triangle, triangle* and star CDW phases under the in-plane lattice constants of 3.399, 3.474 and 3.458 \AA~using different types of exchange-correlation functional including LDA, PBE. The values 3.399 and 3.474 \AA~are the fully relaxed lattice constants using LDA and PBE, respectively. The value 3.458 is the  in-plane lattice constant of bulk $2H$-NbSe$_2$ reported in the reference~\cite{Malliakas2013}. The related total energies are listed in Table \ref{tabs:str_pho}.  It is found  that all the three CDW phases exist regardless of the lattice constants and exchangle-correlation functional being used. Furthermore, it is reasonable that the total energies of the three CDW phases are found to be lower than that of the non-CDW phase. In detail, the triangle and triangle* phase exhibit similar total energies, which are slightly lower than that of the star phase. In particular, the related total energies of the triangle, triangle* and star phase are 0, 0.3 and 2.4 meV per chemical formula respectively at the lattice constant $a = 3.474$~\AA.

\section{The STM  topography}

\begin{figure}[H]
	\centering
	\includegraphics[width=150mm]{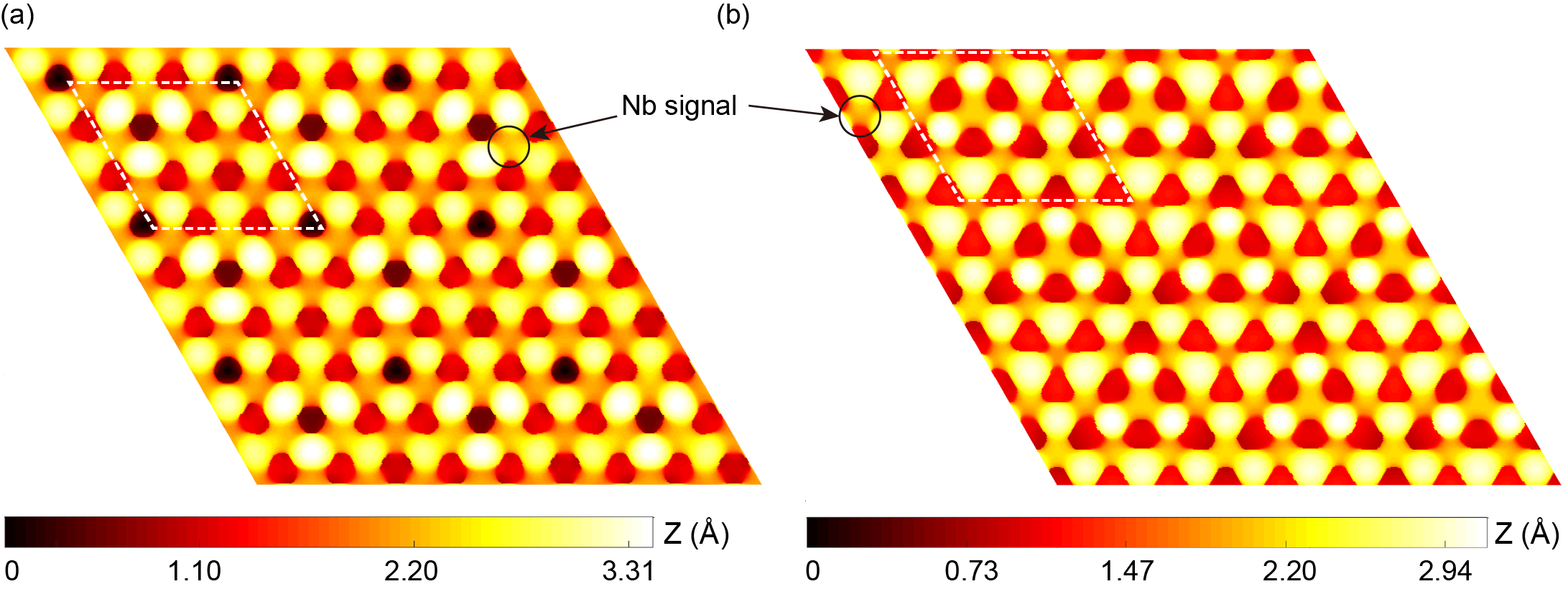} \\
	\caption{ Computed STM topography of monolayer NbSe$_2$ in the triangle (a) and star (b) CDW phases. The colors in the map represent the height function $z(\mathbf{x}, \mathbf{y})$ as mentioned in the main text, where $\mathbf{x}$  and $\mathbf{y}$ are in-plane coordinates. The height of Nb layer was set to be $z$ = 0.}
	\label{figs:STM2}
\end{figure}

In the calculations of the STM topographies for the CDW phases, non-self-consistent calculations was carried out with a  denser $\mathbf{k}$-grid 48$\times$48 after self-consistent calculations. Fig.~\ref{figs:STM2} displays the computed STM topography of monolayer NbSe$_2$ in the triangle and star CDW phases. They are similar to the STM topography displayed in Figs. 2(a) and 2(b) in the main text but with different colorbars. By comparing the STM signals within the 3$\times$3 CDW supercell (bounded by white dashed lines) in Figs.~\ref{figs:STM2}(a) and \ref{figs:STM2}(b) to the atomic positions  within the corresponding supercells (bounded by black dashed lines) in Figs. 2(c) and 2(d) in the main text, one can find that the STM signal is dominated by Se atoms.

\section{Details of spin spiral calculations}

\begin{figure}[H]
	\centering
	\includegraphics[width=150mm]{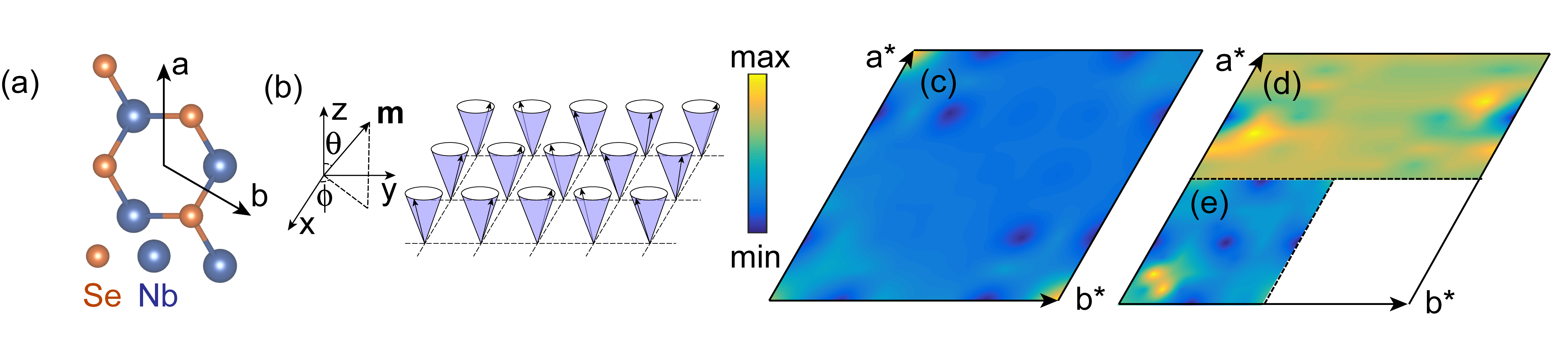} \\
	\caption{(a) $2\times2$ supercell of monolayer NbSe$_2$. \textbf{a} and \textbf{b} are lattice vectors of primitive cell. (b) A schematic diagram of spin spiral structures. (c)-(e): Spin spiral calculated total energy with respect to spin spiral vectors in $1\times1$, $2\times1$, and $2\times2$ supercells, respectively. Here \textbf{a}* and \textbf{b}* are reciprocal lattice vectors of primitive cell. For supercells, the total energy with respect to spin spiral vectors are drawn in corresponding reciprocal cell of their own primitive lattice. The relationship of size between reciprocal cells of different supercells is depicted as well. 
	}
	\label{fig:spiral}
\end{figure}

In order to examine the magnetic ground state(s) suggested by the spin susceptibility calculations, spin spiral calculations were carried out to determine the magnetic order of monolayer NbSe$_2$ in non-CDW phase. As depicted in Fig.~\ref{fig:spiral}(b), we shall restrict ourselves to a consideration of spin spiral structures of the form~\cite{Sandratskii1991}
\begin{subequations}
	\begin{align}
	& m_{j,x}=m_{j}\sin\theta\cos\phi \\
	& m_{j,y}=m_{j}\sin\theta\sin\phi \\
	& m_{j,z}=m_{j}\cos\theta \\
	& \phi=\boldsymbol{q}\cdot\left(\boldsymbol{r}_{j}+\boldsymbol{t}_{n}\right)
	\end{align}
\end{subequations}
where $m_{j,x},m_{j,y},m_{j,z}$ denote the three components of atomic magnetic moments, the $\boldsymbol{t}_{n}$ are the lattice translations, the $\boldsymbol{r}_{j}$ are position vectors of atoms within the unit-cell, the $m_{j}$ is the magnetic moment of the $j$th atom, and $\boldsymbol{q}$ is the vector of the spiral order.
Although crystal with spin spiral magnetic order loses its translation symmetry, a generalized Bloch theorem allows us to compute the energy of a supercell with spiral magnetic order using and at the computational cost of a primitive cell. This allows us to quickly and accurately survey the energies for different $\boldsymbol{q}$'s. 

As shown in Fig.~\ref{fig:spiral}(c), the calculated total energy reaches minima at $\boldsymbol{q}=($\textbf{a}*$/4,0)$ (and other equivalent $\boldsymbol{q}$ vectors, \textbf{a}* and \textbf{b}* are reciprocal lattice vectors of primitive cell), indicating a possible magnetic ground state in a $4\times1$ supercell. Subsequently, the spiral magnetic order is imposed on $2\times1$ and $2\times2$ supercells, which allows spin modulation within a supercell. 
Fig.~\ref{fig:spiral}(d) shows the total energy with respect to spiral vectors in $2\times1$ supercell, the minima appear at $\boldsymbol{q}=($\textbf{a}*$/4,0)$ (the reciprocal lattice vector in \textbf{a} direction of $2\times1$ supercell is half of the primitive cell). The results for $2\times2$ supercell are shown in Fig.~\ref{fig:spiral}(e), the spiral vectors for the lowest total energy are $($\textbf{a}*$/4,0)$ as well.
Base on the results of our spin spiral calculations, we infer that a magnetic ground state may appear in a $4\times1$ supercell, which are consistent with the magnetic instability indicated by our previours spin susceptibiity calculations. We then did a magnetic calculation in a $4\times1$ supercell to finally determine the magnetic order. Fig.~3(c) in the main text shows our calculated antiferromagnetic spin configurations, which is the most stable state in the non-CDW phase, in agreement with previous predictions~\cite{Xu2014}.

\section{Unfolding electronic structures }

In the unfolding calculations for the non-CDW and CDW phases of monolayer NbSe$_2$,  3$\times$3 CDW supercells are adapted, each containing 27 atoms in total. The Fermi surfaces are computed on a  96$\times$96 $\mathbf{k}$-grid and subsequently unfolded to the primitive Brillouin zone (PBZ) of the non-CDW phase using unfolding technique~\cite{Allen2013}. In light of this method, the electronic bands in supercell Brillouin zone  are unfolded to PBZ with a weight function $W_{\mathbf{K}J}(\mathbf{G}) = 1/N\sum_{j = 1}^{N}\langle \mathbf{K} J | \hat{T}(\mathbf{r}_j) | \mathbf{K} J \rangle e^{-i(\mathbf{K} + \mathbf{G})\cdot \mathbf{r}_j}$ which measures the degree of the Bloch symmetry belonging to a primitive cell, that an eignstate of a supercell posesses.  The supercell contains $N$ primitive cells.  $| \mathbf{K} J \rangle $ donates an eigenstate of the supercell with  a band index $J$ and a wavevector $\mathbf{K}$.  $\mathbf{G}$ is a reciprocal lattice vector of the supercell, connecting $\mathbf{K}$ to a wavevector $\mathbf{k}$  in the PBZ as $\mathbf{k} = \mathbf{K} + \mathbf{G}$. $\hat{T}(\mathbf{r}_j)$ is the translation corresponding to the $j$th primitive cell in the supercell.

\begin{figure}[H]
	\centering
	\includegraphics[width=150mm]{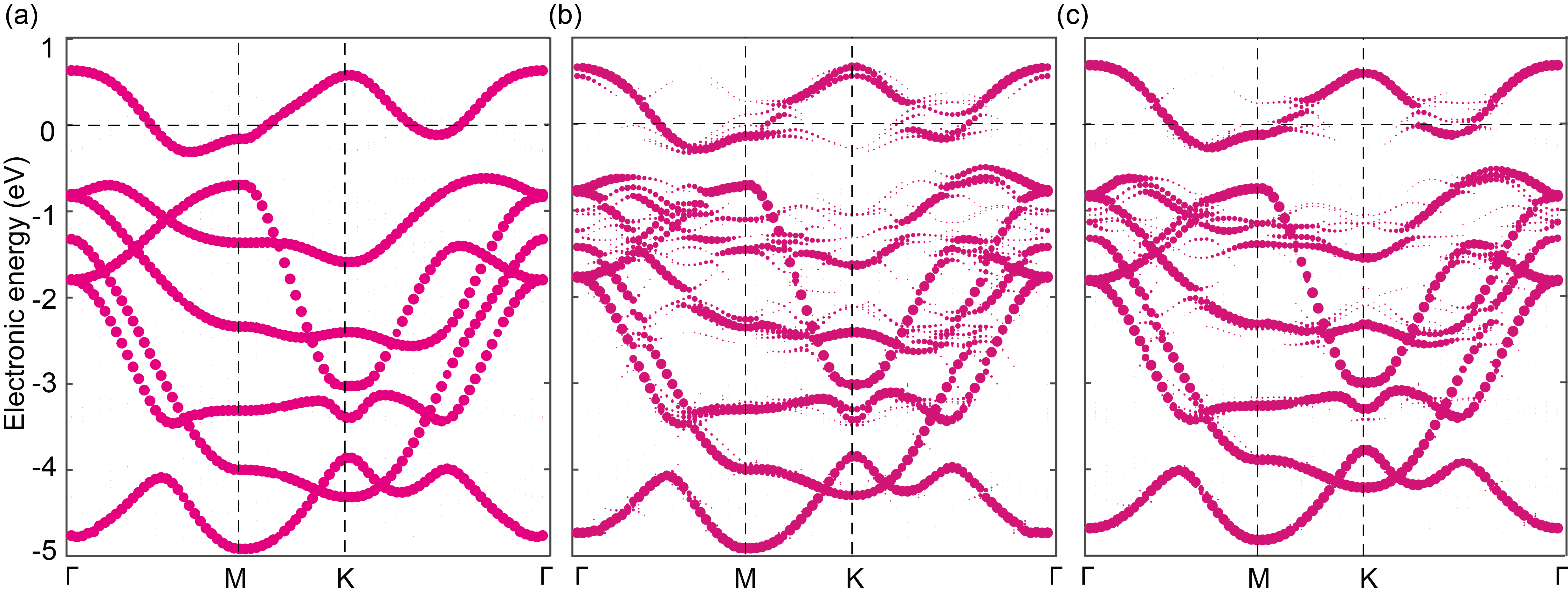} \\
	\caption{Unfolded band structures of the non-CDW phase (a), the triangle CDW phase (b), and the star CDW phase (c), respectively. The sizes of the dots in (a)-(c) denote corresponding unfolding weights $W_{\mathbf{k}J}$. The Fermi levels were aligned. 
	}
	\label{fig:unfold_band}
\end{figure}

Fig.~\ref{fig:unfold_band} displays the electronic band structures of non-CDW phase (a), the triangle CDW phase (b), and the star CDW phase (c) of monolayer NbSe$_2$, respectively. The unfolded band structure of the non-CDW phase agrees perfectly with the one directly computed with a primitive cell. The sizes of the red dots are uniform and the corresponding unfolding weights $W_{\mathbf{k}J}=$ 1.
The most prominent changes of the electronic states after the CDW transition are located in the following three regions:  -2 $\sim$ -2.5 eV, -0.8 $\sim$ -1.6 eV and around the Fermi level. Some parts of the electronic states  are removed compared to that of the non-CDW phase, especially  near the momenta at  2/3$\mathbf{\Gamma}\mathbf{M}$, 1/3$\mathbf{M}\mathbf{K}$, 1/3 and 2/3$\mathbf{K}\mathbf{\Gamma}$. Despite the above changes, the band structures of the  CDW phases coincide with that of the non-CDW phase approximatively, indicating that the shapes of the bandsturcutres remain virtually unchanged and there exist negligible energy shifts for each band in the presence of CDW.

\end{document}